\documentclass[reprint,amsmath,amssymb,aps,twocolumn]{revtex4-2}

\usepackage{graphicx}
\usepackage[caption=false, font=footnotesize]{subfig}
\usepackage[rightcaption]{sidecap}
\usepackage{dcolumn}
\usepackage{bm}
\usepackage{hyperref}
\usepackage{algorithm}
\usepackage{braket} 
\usepackage{tabularx}
\usepackage{graphicx}
\usepackage{xcolor}
\usepackage{rotating}
\usepackage{tikz}
\usetikzlibrary{shapes,snakes}
\usetikzlibrary{calc}
\usetikzlibrary{matrix}
\usepackage{lipsum, babel}
\usepackage[symbol]{footmisc}
\usepackage{float}
\usepackage{physics}
\usepackage{amssymb}
\usepackage[toc,page,titletoc]{appendix}
\usepackage[capitalise,nameinlink]{cleveref}

\crefname{supp}{Supplement}{Supplements}

\newtheorem{definition}{Definition}

\usepackage{amsmath}

\definecolor{pink}{RGB}{237,16,118}
\definecolor{applegreen}{rgb}{0.55, 0.71, 0.0}
\definecolor{celestialblue}{RGB}{62,146,204}
\definecolor{lilanew}{RGB}{152,41,221}

\newcommand{\COMMENT}[1]{}

\newcommand{\GHZ}[1]{$\mathrm{GHZ}_{#1}$}
\newcommand{\key}[1]{%
\hspace*{-7pt}
\begin{turn}{10}
\def \globalscale {0.12}
\begin{tikzpicture}[y=0.80pt, x=-0.80pt, yscale=-\globalscale, xscale=\globalscale, inner sep=0pt, outer sep=0pt]
\begin{scope}[shift={(0,-952.36218)}]
  \path[color=black,fill=black] (23.0000,980.3622) .. controls (32.6339,980.3622) and (40.7318,987.2567) .. (42.5938,996.3622) -- (93.0000,996.3622) .. controls (95.2091,996.3622) and (97.0000,998.1530) .. (97.0000,1000.3622) -- (97.0000,1020.3621) .. controls (97.0000,1022.5711) and (95.2092,1024.3621) .. (93.0000,1024.3621) .. controls (90.7909,1024.3621) and (89.0000,1022.5711) .. (89.0000,1020.3621) -- (89.0000,1004.3622) -- (79.0000,1004.3622) -- (79.0000,1016.3621) .. controls (79.0000,1018.5711) and (77.2092,1020.3621) .. (75.0000,1020.3621) .. controls (72.7909,1020.3621) and (71.0000,1018.5711) .. (71.0000,1016.3621) -- (71.0000,1004.3622) -- (42.5938,1004.3622) .. controls (40.7318,1013.4677) and (32.6339,1020.3622) .. (23.0000,1020.3622) .. controls (12.0017,1020.3622) and (3.0000,1011.3605) .. (3.0000,1000.3622) .. controls (3.0000,989.3640) and (12.0017,980.3622) .. (23.0000,980.3622) -- cycle(23.0000,988.3622) .. controls (16.3253,988.3622) and (11.0000,993.6875) .. (11.0000,1000.3622) .. controls (11.0000,1007.0370) and (16.3253,1012.3622) .. (23.0000,1012.3622) .. controls (29.6747,1012.3622) and (35.0000,1007.0370) .. (35.0000,1000.3622) .. controls (35.0000,993.6875) and (29.6747,988.3622) .. (23.0000,988.3622) -- cycle;
\end{scope}
\end{tikzpicture}
\end{turn}$_{#1}$
\hspace*{-7pt}
}
\newcommand{\ame}[1]{$\bigstar_{#1}$}
\newcommand{\ver}[1]{\checkmark$_{#1}$}
\def\defeq{\mathrel{\mathop:}=} 
\newcommand{\discup}{\mathop{\cup}}

\newcommand{\net}{\mathbf{N}}
\newcommand{\Alice}{A}
\newcommand{\Bobs}{\ahon \setminus \Alice}

\newcommand{\ahon}[0]{\mathbf{P}} 
\newcommand{\uhon}[0]{\mathbf{H}} 
\newcommand{\ucol}[0]{\mathbf{C}}

\newcommand{\hon}[0]{\ahon \discup \uhon}
\newcommand{\col}[0]{\mathbf{C}}

\newcommand{\una}[0]{\mathbf{\bar{P}}}
\newcommand{\aut}[0]{\mathbf{P}}

\newcounter{protocol}
\newenvironment{protocol}[1]
  {\par\addvspace{\topsep}
   \noindent
   \tabularx{\linewidth}{@{} X @{}}
    \hline
    \refstepcounter{protocol}\textbf{Protocol \theprotocol} #1 \\
    \hline}
  { \\
    \hline
   \endtabularx
   \par\addvspace{\topsep}}

\begin{document}
\preprint{APS/123-QED}

\title{Anonymous Quantum Conference Key Agreement}

\author{Frederik Hahn}
\email{frederik.hahn@fu-berlin.de}
 \affiliation{Dahlem Center for Complex Quantum Systems, Freie Universit{\"a}t Berlin, 14195 Berlin, Germany}

\author{Jarn de Jong}%
\affiliation{%
 Electrical Engineering and Computer Science Department, Technische Universit{\"a}t Berlin, 10587 Berlin, Germany}%

\author{Anna Pappa}
 \affiliation{%
 Electrical Engineering and Computer Science Department, Technische Universit{\"a}t Berlin, 10587 Berlin, Germany}%

\date{\today}

\begin{abstract}
Conference Key Agreement (CKA) is a cryptographic effort of multiple parties to establish a shared secret key. In future quantum networks, generating secret keys in an anonymous way is of tremendous importance for parties that want to keep their shared key secret and at the same time protect their own identity. We provide a definition of anonymity for general protocols and present a CKA protocol that is provably anonymous under realistic adversarial scenarios. We base our protocol on shared Greenberger-Horne-Zeilinger states, which have been proposed as more efficient resources for CKA protocols, compared to bipartite entangled resources. The existence of secure and anonymous protocols based on multipartite entangled states provides a new insight on their potential as resources and paves the way for further applications.

\end{abstract}

\maketitle

\section{\label{sec:Introduction}Introduction}
One of the main applications of quantum information processing is to provide additional security for communication. The most common setting is one of two parties, Alice and Bob, who want to establish a shared secret key in order to encrypt further communication. Since their introduction \cite{BB84}, Quantum Key Distribution (QKD) protocols have been proposed and implemented in a standard fashion, although several 
practical challenges remain to be addressed \cite{diamanti_practical_2016}. Here, we examine a more generalised scenario, where several parties want to establish a shared secret key. In this multiparty setting we introduce a new notion of \emph{anonymity}, where we request that the identities of the parties sharing the secret key are all protected. Such scenarios are highly relevant for several reasons.
One example is the case of whistle-blowing; a person might want to broadcast an encrypted message such that specific parties can decrypt it, while keeping the identities of all involved parties secret. For such anonymous whistle-blowing, the underlying protocol needs to involve non-participating parties, such that an authority maintaining the network cannot uncover who takes part in the secret communication.
 To the best of our knowledge,  this is the first multipartite protocol that provides anonymity for a sender and multiple receivers alike.

To succeed in attaining this goal, we need to address two different elements, \emph{anonymity} and \emph{multiparty key generation}. For a concise review of the latter, often referred to as conference key agreement (CKA), we refer the interested reader to \cite{murta_quantum_2020}.
Combining the two elements, we achieve \emph{anonymous conference key agreement}, which allows a sender to transmit a private message to specific receivers of her choice, while keeping their identities secret from external parties and even from each other.

Previous work \cite{christandl_quantum_2005} has shown how to achieve anonymous transmission of classical bits using the correlations natural to the \GHZ{} state \cite{ghz1989} and how to anonymously create bipartite entanglement from a larger \GHZ{} state. In \cite{unnikrishnan_anonymity_2019} the latter is developed further, by adding a scheme for anonymous notification of the receiver and for verification \cite{pappa_multipartite_2012, mccutcheon_experimental_2016} of the anonymous entanglement generation. However, since extracting multiple bipartite Bell states from a single \GHZ{} state is impossible, we need an alternative approach that enables us to perform anonymous CKA between a subset of a given network. One approach could be to use other multipartite entangled quantum states  \cite{leung_quantum_2010, hahn_quantum_2019, PhysRevA.92.032316} to create bipartite entanglement between the sender and all receivers separately; however, that would increase the use of quantum resources. We show that it is in fact possible to anonymously establish the necessary entanglement between sender and receivers simultaneously, using a single \GHZ{} state shared by a source through the network.

In this paper, we introduce a protocol to establish a secret key between the sender `Alice' and $m$ receiving parties of her choice. We use both `Bob' and `receiver' to refer to each of those receiving parties and `participants' to refer to Alice and all Bobs. 
The $m+1 \leq n$ participants are part of a larger network of $n$ parties.
The $m$ Bobs are notified anonymously by Alice through a notification protocol. 
A large \GHZ{} state $\frac{1}{\sqrt{2}}\left(\ket{0}^n+\ket{1}^n\right)$ is then shared between the $n$ parties, which can either be done centrally or using a given network infrastructure via quantum repeaters or quantum network coding \cite{epping_multi-partite_2017}. From this \GHZ{n} state, we subsequently show how to anonymously extract a \GHZ{m+1} state shared only between the participants. The resulting state can be either verified or used to run the CKA protocol. Both the participants' identities and their shared key are hidden from an attacker `Eve' in our protocols. We either assume Eve to follow the protocol and control a single node in the network, or to diverge from the protocol and control multiple non-participating nodes.

\section{Preliminaries}
\noindent We label with $\net$ the set of all $n\defeq \abs{\net}$ parties in the network and with $\aut \defeq \{\Alice,B_{1},\ldots,B_{m}\}$ the set of the protocol's participants, where $\Alice$ refers to Alice and $\{B_{i}\}$ to the $m$ Bobs chosen by her. 

Let Eve be an attacker whose goal it is to learn $\aut$. If Eve corrupts some parties, she trivially learns their role in the protocol, i.e.~whether or not they belong to $\aut$. By $\mathcal{I}_{\mathrm{Eve}}$ we denote this information as well as any prior information on $\left\{\Pr(\mathbf{G}=\ahon)\right\}_{\mathbf{G} \subset \net}$, i.e.~the probability distribution that a subset $\mathbf{G}$ of the parties is equal to $\ahon$. Denoting with $\mathcal{I}^{+}_{\mathrm{Eve}}$ the \textit{additional} information that becomes available to Eve during the protocol, we can define \textit{anonymity} by demanding that running the protocol increases Eve's knowledge only in a trivial way.
\begin{definition}[Anonymity]\label{def:anonymity}
	A protocol is \textit{anonymous} from the perspective of Eve if for all subsets $\mathbf{G} \subset \net$
	\begin{equation}
	\Pr\left(\mathbf{G} =\ahon \mid \mathcal{I}_{\mathrm{Eve}}^+,\mathcal{I}_{\mathrm{Eve}}\right) = \Pr(\mathbf{G} =\ahon \mid \mathcal{I}_{\mathrm{Eve}}),
	\label{eq:privacy}
	\end{equation}
	where $\mathcal{I}_{\mathrm{Eve}}^{+}$ is the information that becomes available to Eve during the protocol and $\mathcal{I}_{\mathrm{Eve}}$ is both the information that Eve has beforehand and trivial information that she obtains about the parties that she corrupts.  
\end{definition}

\noindent Here, by trivial information we mean the information that is available to each party regarding their role in the protocol, i.e.~whether they belong in $\aut$ or not. In the context of key agreement, we can assume that the participants are not corrupted by a fully malicious Eve, since this would jeopardise the whole key. We therefore assume that they are \emph{honest-but-curious}, i.e.~that they obey the protocol in order to establish a key, but may otherwise be interested in learning other participants' identities. For the non-participating parties we consider the same honest-but-curious model, as well as a fully dishonest one. Hence, $\net$ can be partitioned into the three disjoint sets of:
\begin{description}
\item[$\ahon$] honest-but-curious participating parties,
\item[$\uhon$] honest-but-curious non-participating parties,
\item[$\col$] dishonest and colluding non-participating parties.
\end{description}

We either assume Eve to follow the protocol and control a single party in $\ahon$ or $\uhon$, or to diverge from the protocol and control $\col$. Note however that our definition of anonymity is applicable to other corruption models and therefore applies more generally to any cryptographic protocol. 

As previously mentioned, our CKA protocol exploits the correlations of a shared GHZ state to generate the conference key. Since the parties in $\col$ could apply an arbitrary quantum map to their system, this would result in a state  $\epsilon$-close to $\rho_{\net} \defeq \ketbra{\net}{\net}$, with $\ket{\net}$ equal to
\begin{equation}
\frac{1}{\sqrt{2}}\left(\ket{0 \hdots 0}_{\hon} \otimes \ket{\Psi}_{\col} + \ket{1\hdots 1}_{\hon} \otimes \ket{\Phi}_{\col}\right).
\label{eq:GHZ}
\end{equation}
Here, the two states on $\col$ need not be orthogonal. They neither need to be pure, but since mixed states do not offer an advantage to Eve we may assume they are. For a discussion on untrusted or faulty sources we refer to the Discussion.

With the above definitions, we are now ready to introduce the subprotocols of the Anonymous Conference Key Agreement protocol. All protocols we propose are anonymous according to Def.~\ref{def:anonymity}, with the corresponding proofs detailed in the Appendix.

\section{Generating Anonymous Multiparty Entanglement}

\begin{figure*}
 \centering
 \includegraphics[width=\linewidth]{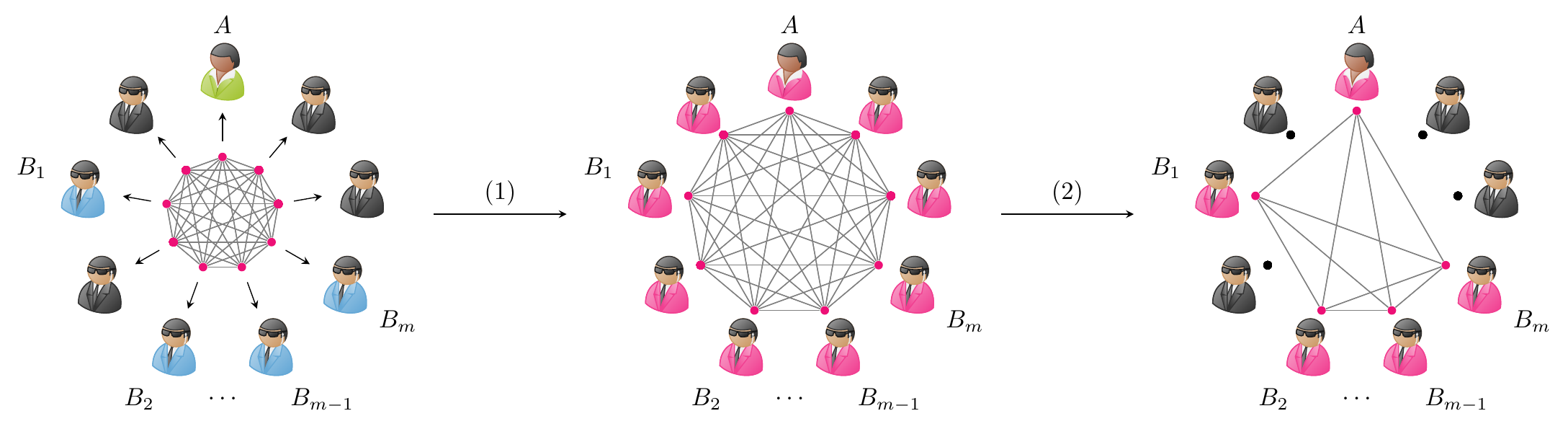}
\caption{Visualisation of Protocol~\ref{AME}. A \GHZ{n} state is shared with all agents left of arrow $(1)$. Here, the participants are highlighted in \textcolor{applegreen}{green} and \textcolor{celestialblue}{blue}. tSince the shared \GHZ{n} state is agnostic of the receivers' identities and all agents are entangled right of arrow $(1)$, they are all highlighted in \textcolor{pink}{pink}. Right of arrow $(2)$, all non-participating parties are disentangled and therefore not highlighted anymore. The $m$ Bobs and Alice now share a \GHZ{m+1} state after completing the steps of \texttt{AME}.}
\label{fig:AME}
\end{figure*}

We start by presenting two sub-protocols, namely \texttt{Notification} and \texttt{Anonymous Multiparty Entanglement} \texttt{(AME)}. Our version of \texttt{Notification} is based on \cite{broadbent_information-theoretic_2007} and is a classical protocol used by Alice to notify the $m$ receiving agents, while maintaining anonymity for all parties involved. The protocol requires pairwise private classical communication -- which can be established using a key generation protocol with a Bell pair -- and access to  private sources of randomness. An illustration of Protocol~\ref{Notification} can be found in App.~\ref{App:Notification}.
\vspace{0.1in}

\begin{protocol}{\label{Notification} \texttt{Notification}}
\textit{Input.} Alice's choice of $m$ receivers.\\
\textit{Goal.} The $m$ receivers get notified.

\vspace{\baselineskip}
\noindent For agent $i=1,\dots,n$:
\begin{enumerate}
    \item All agents $j\in\{1,\ldots,n\}$ do the following.
    \begin{enumerate}
        \item When $j$ corresponds to Alice ($j_a$), and $i$ is not a receiver, she chooses $n$ random bits $\{r_{j,k}^i\}_{k=1}^{n}$ such that $\bigoplus^n_{k=1} r_{j,k}^i = 0$. If $i$ is a receiver, she chooses $n$ random bits such that $\bigoplus^n_{k=1} r_{j,k}^i = 1$. She sends bit $r_{j,k}^i$ to agent $k$.
        \item When $j\neq j_a$, the agent chooses $n$ random bits $\{r_{j,k}^i\}_{k=1}^{n}$ such that $\bigoplus^n_{k=1} r_{j,k}^i = 0$ and sends bit $r_{j,k}^i$ to agent $k$.
    \end{enumerate}
    \item All agents $k \in \{1,\ldots,n\}$ receive $\{r_{j,k}^{i}\}_{j=1}^{n}$ , compute $z_k^i=\bigoplus_{j=1}^n r_{j,k}^i$ and send it to agent $i$.
    \item Agent $i$ takes the received $\{z_{k}^{i}\}_{k=1}^{n}$ to compute $z^i=\bigoplus_{k=1}^n z_k^i$; if $z^i=1$ they are thereby notified to be a designated receiver.
\end{enumerate}
\end{protocol}\label{Notify}

\noindent{\bf Analysis:} Anonymity is maintained following the work of \cite{broadbent_information-theoretic_2007}. Remember that by the nature of our goal, the identities of the Bobs are available to Alice since she has chosen them. The \texttt{Notification} protocol requires $\mathcal{O}(n^3)$ communication channel uses between pairs of parties. Note that the \texttt{Notification} protocol is  allowing Alice to anonymously transmit the same bit to all receivers to establish a common key. Such a process would however be extremely inefficient; if one Bell pair is required for each private classical communication round, then for each bit of generated key, $\mathcal{O}(n^3)$ Bell pairs would be consumed. If instead we use \texttt{Notification} only once to notify the receivers, we can exploit the properties of the  shared multipartite entanglement to establish a common key more efficiently while maintaining the anonymity that Protocol~\ref{Notification} provides.

We now introduce the second subprotocol \texttt{AME}, visualised in Fig.~\ref{fig:AME}. As a generalisation of the protocol first proposed in \cite{christandl_quantum_2005} for anonymously distributing Bell states, it is a protocol for anonymously establishing \GHZ{} states. Here, $n$ parties are sharing a \GHZ{} state, and $m+1$ of them (Alice and $m$ receivers) want to anonymously end up with a smaller, $(m+1)$-partite \GHZ{} state. To achieve this, all parties require access to a broadcast channel -- a necessary requirement to achieve any type of anonymity for participants in a communication setting \cite{fitzi_unconditional_2002}.

\begin{protocol}{ \texttt{Anonymous Multiparty Entanglement}\label{AME}}
\textit{Input.} A shared \GHZ{n} state; Alice knowing the identities of the non-participants $\una$. \\
\textit{Goal.} A \GHZ{m+1} state shared between $\aut$.
\begin{enumerate}
  \item Alice and the Bobs each draw a random bit. Everyone else 
  measures in the $X$-basis, yielding a measurement outcome bit $x_{i}$ for $i\in \una$.\label{AME_step1}
\item All parties broadcast their bits in a random order or, if possible, simultaneously.\label{AME_step2}
\item Alice applies a $Z$ gate if the parity of the non-participating parties' bits is odd.\label{AME_step3}
  \end{enumerate}
  \end{protocol}

\noindent{\bf Analysis:} The correctness of the protocol follows from the proof in \cite{christandl_quantum_2005}. With the Hadamard matrix $H$ we can rewrite the \GHZ{n} state as proportional to

\begin{equation*}
\sum_{x\in\{0,1\}^{\abs{\una}}}\left(\ket{0 \hdots 0}_{\aut}+(-1)^{\Delta(x)}\ket{1 \hdots 1}_{\aut}\right)\otimes H_{\una}\ket{x}_{\una},
\end{equation*}
where $\Delta(x)$ is the Hamming weight of $x$ and the subscripts $\aut$ and $\una$ indicate the participating and non-participating parties, respectively.
Since $H$ interchanges the $X$- and $Z$-bases, the state shared between Alice and the Bobs after the $X$-measurements of Step~\ref{AME_step1} is  $\frac{1}{\sqrt{2}}\big(\ket{0 \hdots 0}_{\aut}+(-1)^{\Delta(x)}\ket{1 \hdots 1}_{\aut}\big)$, where $x$ contains all measurement outcomes announced in Step~\ref{AME_step2}. Finally, calculating $\Delta(x)$ in Step~\ref{AME_step3}, Alice locally corrects the state to obtain the desired \GHZ{m+1} state.

With respect to anonymity, the key elements are the intrinsic correlations of \GHZ{} states. As observed in \cite{christandl_quantum_2005}, any rotation around the $\hat{z}$-axis applied to any qubit of a \GHZ{} state has the same effect on the global state independent of the chosen qubit. To correct the state, Alice only needs the parity of the measurement outcomes of the non-participating parties, yet, masking their identity, each Bob announces a random bit too. No information about the operations performed by the different parties can be inferred, since all announced bits can be shown to be uniformly random and a $Z$-gate does not reveal the position of the qubit it was applied to either. Only Alice knows the identities of the Bobs, so only she is able to discern the measurement outcomes from the random bits. For a detailed discussion on why the protocol does not leak any information about the identity of either Alice or the Bobs in untrusted settings, we refer to App.~\ref{appendix:proofsofanonymity}. 

A combination of the above two protocols allows for an anonymous distribution of a \GHZ{m+1} state, which in turn can be measured in the $Z$-basis by all participants to generate a shared secret key. However, to be secure against dishonest or eavesdropping parties, the state needs to be verified.

\section{Anonymous Quantum Conference Key Agreement}\label{sec:Verificationprotocol}

\noindent In the setting of an untrusted source any verification could be performed immediately after the distribution of the state. However, a party in $\una$ might not measure in Protocol~\ref{AME}, and thereby be part of the extracted, then $(>m+1)$-partite, \GHZ{} state. This security risk was independently noticed in \cite{yang_examining_2020} for the case of two-party communication. To detect both a faulty source and dishonest parties, the verification of the state has to be postponed until \textit{after} Protocol \ref{AME}. 
Note that in this setting, only the communication of authorized parties will be considered by Alice. 
Protocol~\ref{Ver_GHZ} verifies that the state on $\ahon$ is close to the \GHZ{m+1} state, and therefore also disentangled from all other parties, including $\col$. Protocol~\ref{Ver_GHZ} is similar to \cite{pappa_multipartite_2012} and inspired by the pseudotelepathy studies of \cite{brassard_multi-party_2003}, but adjusted here to protect the identities of the participants and to always set the verifier to be Alice. It requires private sources of randomness and a classical broadcasting channel.
\vspace{0.1in}

\begin{protocol}{ \texttt{Verification}\label{Ver_GHZ}}
\textit{Input.} A shared state between $\abs{\ahon}=m+1$ parties.
\newline
\textit{Goal.} Verification or rejection of the shared state as a \GHZ{m+1} state by Alice.
\newline
\begin{enumerate}
  \item Every $B_{i}$ draws a random bit $b_{i}$ and measures in the $X$- or $Y$-basis if it equals $0$ or $1$ respectively, obtaining a measurement outcome $o_{i}$.\label{VER_step2}
  \item Everyone broadcasts $(b_{i},o_{i})$, including Alice, who chooses her bits $(b_{0},o_{0})$ at random.
  \item Alice resets her bit such that $\sum_{i = 0}^{m} b_{i}=0 \pmod 2$. She measures in the $X$- or $Y$-basis if her bit equals $0$ or $1$ respectively, thereby also resetting $o_{0}$.\label{VER_step4}
  \item If and only if $\frac{1}{2}\sum_{i} b_{i}+\sum_{i = 0}^{m} o_{i}   = 0\pmod{2}$, Alice accepts the state.

\end{enumerate}
\end{protocol}

\noindent{\bf Analysis:}
From \cite{pappa_multipartite_2012} we know that the state is verified to be increasingly close to the \GHZ{} state with the number of passed \texttt{Verification} rounds. To mask their identity, the parties in  $\ahon$ need both $\uhon$ and $\col$ to announce random bits as well. This renders all public communication uniformly random. Since the relevant quantum correlations are only accessible to Alice, all parties are indistinguishable from the perspective of Eve.  We refer to  App.~\ref{appendix:proofsofanonymity} for further details. 

We are now ready to define Protocol~\ref{AKA_ver1} for anonymously sharing a key between $\aut$, where we introduce the parameters $L$ as the number of shared GHZ-states and $D$ as a parameter both determining the level of security and the length of the generated shared key. The main difference between the proposed protocol and the one in \cite{unnikrishnan_anonymity_2019} is that the non-participating parties are asked to announce random values to mask the identities of the authorized parties and that the protocol aborts if the values are not announced in time. Protocol \ref{AKA_ver1} combines all previous protocols and additionally requires a public source of randomness.

\begin{protocol}{ \texttt{Anonymous Conference Key Agreement}\label{AKA_ver1}}
\textit{Input.} Alice as initiator; parameters $L$ and $D$. 
\newline
\textit{Goal.} Anonymous generation of secret key between $\aut$.
\newline
\begin{enumerate}
  \item Alice notifies the $m$ Bobs by running the \texttt{Notification} protocol.\label{AVKA_Notification}
  \item The source generates and shares $L$ \GHZ{} states.\label{AVKA_L_GHZ_states}
  \item The parties run the \texttt{AME} protocol on them.
  \item The parties ask a public source of randomness to broadcast a bit $b$ such that $\Pr[b=1]=\frac{1}{D}$.
  \begin{description}
   \item[\texttt{Verification} round] If $b=0$, Alice runs the \texttt{Verification} protocol on the $(m+1)$-partite state. The remaining parties announce random values. 
  \item[\texttt{KeyGen} round] If $b=1$, Alice and the Bobs $Z$-measure to obtain a shared secret bit.
\end{description}
\item If Alice is content with the checks of the \texttt{Verification} protocol, she can anonymously validate the protocol.
\end{enumerate}
\end{protocol}

\noindent{\bf Analysis:} The above protocol establishes a secret key between the participants, while keeping their identities secret from both  outsiders and each other. The \texttt{Veri}\-\texttt{fication} rounds ensure that the state on  $\aut$ is $\epsilon$-close to the \GHZ{m+1}, which exhibits correlations that only Alice can observe. Likewise, neither the public communication nor the remainder of the state
are correlated with the identities. On average $D-1$ out of $D$ states will be used to verify the state and only one to provide a secret key; therefore the key rate of Protocol~\ref{AKA_ver1} approaches  $\frac{L}{D}$ in the asymptotic regime. We refer to  App.~\ref{appendix:proofsofanonymity}  for a detailed proof of anonymity and to the Discussion for the case where Alice does \textit{not} accept the shared state. 

Note that \texttt{Verification} implicitly verifies the \texttt{Notification} protocol, as the bits that Alice takes into consideration will not have the correct correlations otherwise. 
It is further worth mentioning that as presented, all protocols are self-contained. However, when combined, one could reduce both the communication overhead and the number of applied quantum operations. Specifically, instead of outputting random values, the participants could simply announce the outputs of the verification process during the next round. In the same spirit, Alice does not need to perform the $Z$-correction at the end of the \texttt{AME} protocol, since she can choose a complementary  set of stabiliser measurements during the \texttt{Verification} protocol.

\section{Discussion}\label{sec:discussion}
We demonstrated how to efficiently achieve anonymity for conference key agreement by using multipartite quantum states. Starting from a large \GHZ{} state shared between $n$ parties, our method enables a sender to anonymously notify a set of receivers and establish a secret key.
While here we focused on \GHZ{} states, other types of quantum states have also been used for creating anonymous entanglement, as well as for CKA \cite{lipinska_anonymous_2018,grasselli_conference_2019}; it is however unknown whether we can combine these to achieve the same task as presented here.

We assumed that the source is not actively malicious; the protocol will abort if the state is not close to the \GHZ{} state, 
but anonymity is then not guaranteed. The \texttt{AME} protocol is run before each \texttt{Verification} round, which means that a privacy leak during the \texttt{AME} round due to an actively malicious source can never be caught in time. This is easily fixed by additionally verifying the \GHZ{n} after its initial sharing but omitted here for simplicity. We note however that an anonymous version of the protocol in \cite{pappa_multipartite_2012} should be performed, similar to Protocol \ref{Ver_GHZ}.

Finally, practical sources and channels can be faulty and hence the need for anonymous error correction and privacy amplification arises \cite{epping_multi-partite_2017, proietti_experimental_2020}. 
We intend to address this in follow-up work, together with the finite-key effects of real-world implementations.

\noindent{} {\bf Acknowledgments:} A.P.~and J.d.J.~acknowledge support from the DFG via the Emmy Noether grant 418294583 and F.H.~ from the Studienstiftung des deutschen Volkes. We want to thank Stefanie Barz, Julius Wallnöfer and Nathan Walk for useful discussions.

\clearpage
\onecolumngrid 

\appendix
\section{Visualisation of \texttt{Notification}}\label{App:Notification}

\begin{figure*}[b!]
 \centering
 \subfloat[\centering \label{fig:NotificationA} Step~\ref{Notification_step1a} with $j_a=1$.]{\includegraphics[width=0.23\linewidth]{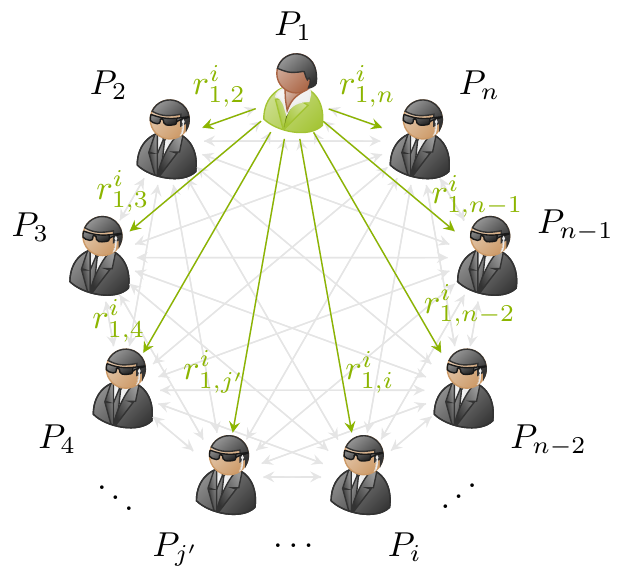}}
 \hfill
 \subfloat[\centering \label{fig:NotificationB} Step~\ref{Notification_step1b} with $j=j'$.]{\includegraphics[width=0.23\linewidth]{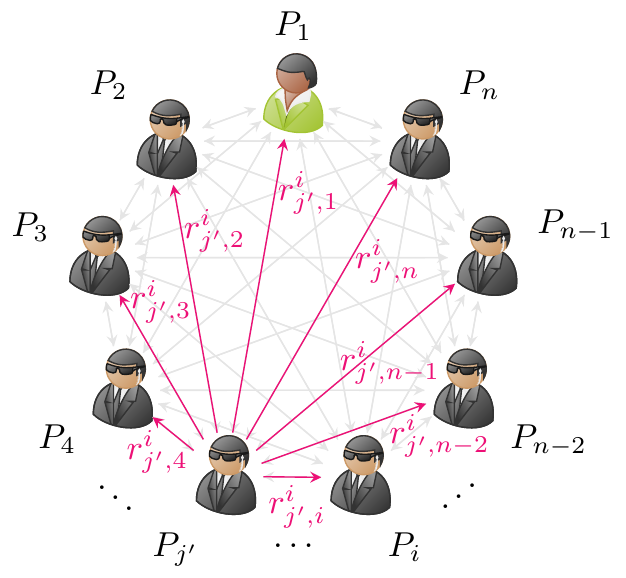}}
 \hfill
 \subfloat[\centering \label{fig:NotificationC} Step~\ref{Notification_step2} with $k=j'$]{\includegraphics[width=0.23\linewidth]{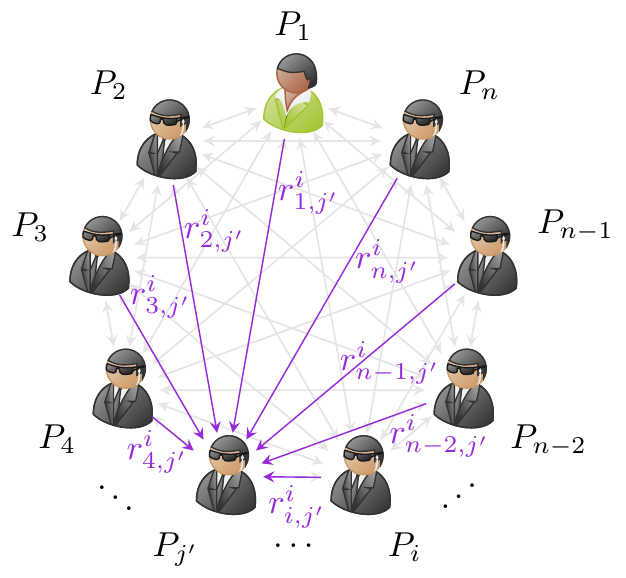}}
  \subfloat[\centering \label{fig:NotificationD} Step~\ref{Notification_step3} with $i=i_b$.]{\includegraphics[width=0.23\linewidth]{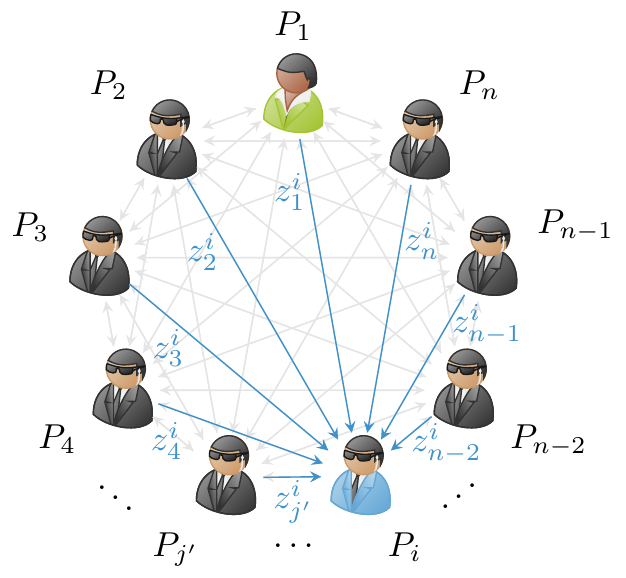}}
    \\
    \vspace{\baselineskip}
    \includegraphics[width=0.6\textwidth]{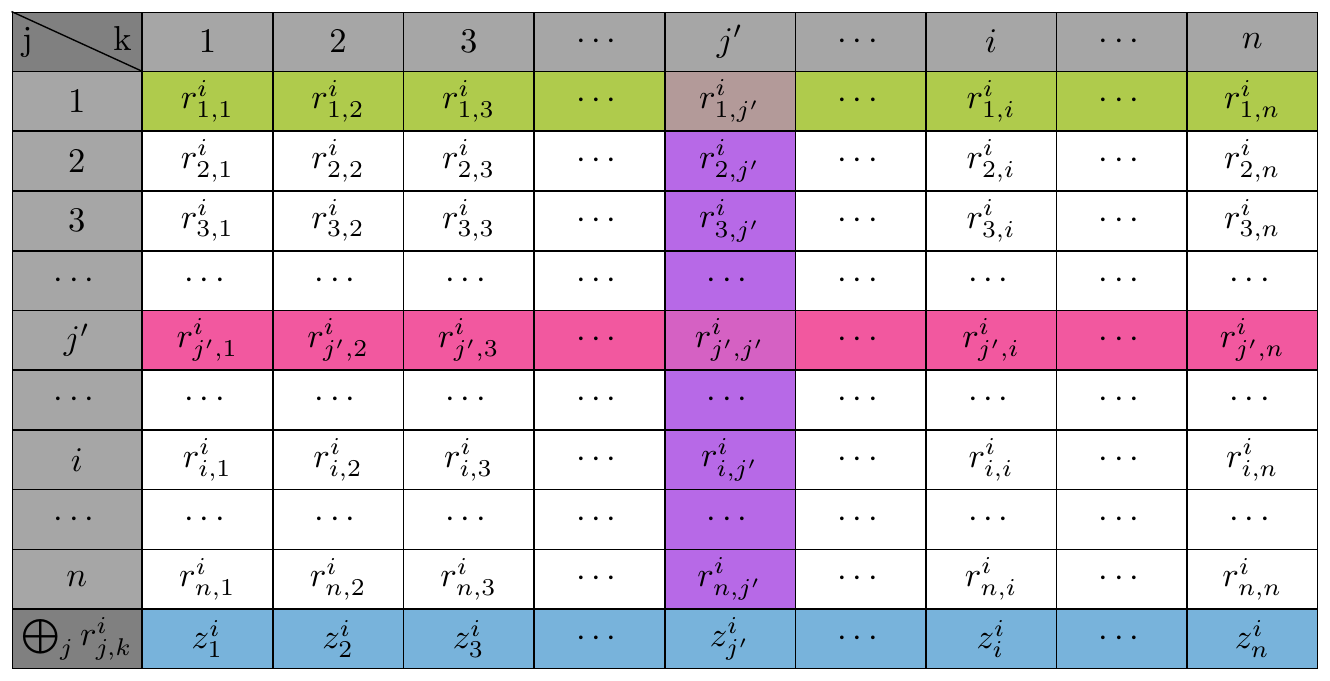}
    \caption{Visualisation of Protocol~\ref{Notification}. The table contains all $r^{i}_{j,k}$ for a fixed agent $P_i\in\net$ in the \texttt{Notification} protocol. Here, we identify Alice with $P_1$. She chooses $\{r^{i}_{1,k}\}_{k=1}^n$ and sends them to $P_k$ in Step~\ref{Notification_step1a} (\textcolor{applegreen}{Fig.~\ref{fig:NotificationA}}). Note that only if $P_i$ is a receiver, the \textcolor{applegreen}{green} row adds up to $1 \pmod 2$; otherwise to $0 \pmod 2$. 
	Analogously, the \textcolor{pink}{pink} highlighting shows Step~\ref{Notification_step1b} from the perspective of $P_{j'}$ (\textcolor{pink}{Fig.~\ref{fig:NotificationB}}). This and all other rows add up to $0 \pmod 2$.
	The $\{r^{i}_{j,j'}\}_{j=1}^{n}$ that $P_{j'}$ receives in Step~\ref{Notification_step2}  (\textcolor{lilanew}{Fig.~\ref{fig:NotificationC}}) are highlighted in \textcolor{lilanew}{purple}.   The last row, highlighted in \textcolor{celestialblue}{blue}, shows the $\{z^{i}_{k}\}_{k=1}^{n}$ received by $P_i$ in Step~\ref{Notification_step3} (\textcolor{celestialblue}{Fig.~\ref{fig:NotificationD}}).  By construction, only if $P_i$  is a receiver, it adds up to $1 \pmod 2$.}
	\label{fig:Notification}
\end{figure*}

\setcounter{protocol}{0}
\begin{protocol}{\label{Notification} \texttt{Notification}}
\textit{Input.} Alice's choice of $m$ receivers.\\
\textit{Goal.} The $m$ receivers get notified.

\vspace{\baselineskip}
\noindent For agent $i=1,\dots,n$:
\begin{enumerate}
    \item All agents $j\in\{1,\ldots,n\}$ do the following.\label{Notification_step1}
    \begin{enumerate}
        \item When $j$ corresponds to Alice ($j_a$), and $i$ is not a receiver, she chooses $n$ random bits $\{r_{j,k}^i\}_{k=1}^{n}$ such that $\bigoplus^n_{k=1} r_{j,k}^i = 0$. If $i$ is a receiver, she chooses $n$ random bits such that $\bigoplus^n_{k=1} r_{j,k}^i = 1$. She sends bit $r_{j,k}^i$ to agent $k$ (\textcolor{applegreen}{Fig.~\ref{fig:NotificationA}}).\label{Notification_step1a}
        \item When $j\neq j_a$, the agent chooses $n$ random bits $\{r_{j,k}^i\}_{k=1}^{n}$ such that $\bigoplus^n_{k=1} r_{j,k}^i = 0$ and sends bit $r_{j,k}^i$ to agent $k$ (\textcolor{pink}{Fig.~\ref{fig:NotificationB}}).\label{Notification_step1b}
    \end{enumerate}
    \item All agents $k \in \{1,\ldots,n\}$ receive $\{r_{j,k}^{i}\}_{j=1}^{n}$ (\textcolor{lilanew}{Fig.~\ref{fig:NotificationC}}), compute $z_k^i=\bigoplus_{j=1}^n r_{j,k}^i$ and send it to agent $i$.\label{Notification_step2}
    \item Agent $i$ takes the received $\{z_{k}^{i}\}_{k=1}^{n}$ (\textcolor{celestialblue}{Fig. ~\ref{fig:NotificationD}}) to compute $z^i=\bigoplus_{k=1}^n z_k^i$; if $z^i=1$ they are thereby notified to be a designated receiver.\label{Notification_step3}
\end{enumerate}
\end{protocol}

\clearpage

\section{Anonymity of the Conference Key Agreement}
\label{appendix:proofsofanonymity}
\setcounter{definition}{0}
\begin{definition}[Anonymity]\label{def:anonymity}
	A protocol is \textit{anonymous} from the perspective of Eve if for all subsets $\mathbf{G} \subset \net$
	\begin{equation}
	\Pr\left(\mathbf{G} =\ahon \mid \mathcal{I}_{\mathrm{Eve}}^+,\mathcal{I}_{\mathrm{Eve}}\right) = \Pr(\mathbf{G} =\ahon \mid \mathcal{I}_{\mathrm{Eve}}),
	\label{eq:privacy}
	\end{equation}
	where $\mathcal{I}_{\mathrm{Eve}}^{+}$ is the information that becomes available to Eve during the protocol and $\mathcal{I}_{\mathrm{Eve}}$ is both the information that Eve has beforehand and trivial information that she obtains about the parties that she corrupts.  
\end{definition}
In order to satisfy Eq.~\eqref{eq:privacy}, $\mathcal{I}^{+}_{\mathrm{Eve}}$
should not change Eve's probability distribution of uncovering
the partitioning of $\net$ into its constituents; it 
does not reveal anything about $\aut$, $\uhon$ or -- implicitly -- about $\col$.
Apart from the trivial attacker $\Alice$ we consider three different types of Eve, namely any party in $\Bobs$ or $\uhon$ or all parties in $\col$.
As a shorthand for our subprotocols, we introduce the symbols \ame{}  (\texttt{AME}), \ver{}  (\texttt{Verification}) and \key{} (\texttt{KeyGen}).

\begin{table*}[h!]
\includegraphics[height=4.5cm
]{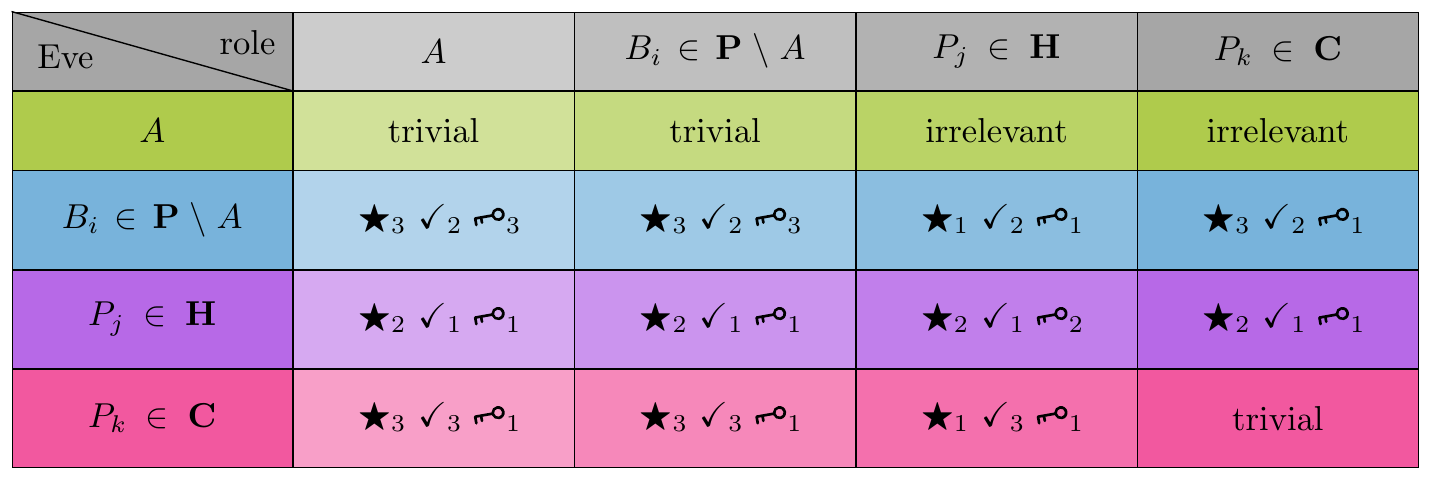}
  \caption{The rows are labeled by the types of Eve and the columns by the roles that Eve may try to uncover. The first row is mostly trivial, since the protocol is designed such that $\Alice$ chooses the partitioning $\net=\aut\discup\una$ herself and it is irrelevant that she is unaware of who in $\una$ is colluding. The arguments corresponding to the symbols are given in Sec.~\ref{app:AME}, \ref{app:Ver} and \ref{app:key}.
  }
  \label{tab:perspectivefrom}
\end{table*}

\noindent We use the structure of Tab.~\ref{tab:perspectivefrom} to prove anonymity with respect to all different types of Eve. 
For the \texttt{Notification} protocol we refer to the original paper of Broadbent and Tapp.
The \texttt{AME} protocol and the \texttt{Verification} protocol will be examined in Sec.~\ref{app:AME} and \ref{app:Ver}. The \texttt{KeyGen} subprotocol does not require any public communication and will be examined in Sec.~\ref{app:key}. 
To prove our claim we consider the following two aspects.  The \emph{public communication} (cf.~Tab.~\ref{tab:publiccommunication}) throughout the protocol does not help Eve to reveal the roles of the participating parties. We prove this by showing that all public communication is indistinguishable from Eve's point of view. 
As $\Alice$ announces only uniformly random and uncorrelated bits, we will show the same for the parties in $\Bobs$, $\uhon$ and $\col$ from Eve's perspective.  Likewise, the \emph{quantum states} accessible to Eve do not help her to reveal the roles of the participating parties, even given access to the public communication. This means that the post-measurement states of Eve can neither be correlated with the measurement outcomes of other parties, nor with any direct information regarding their roles. Note that the global quantum state may encode such information regarding the roles as long as it is not accessible to anyone but Alice.

\begin{table*}[h!]
 \centering
\includegraphics[height=4.5cm
]{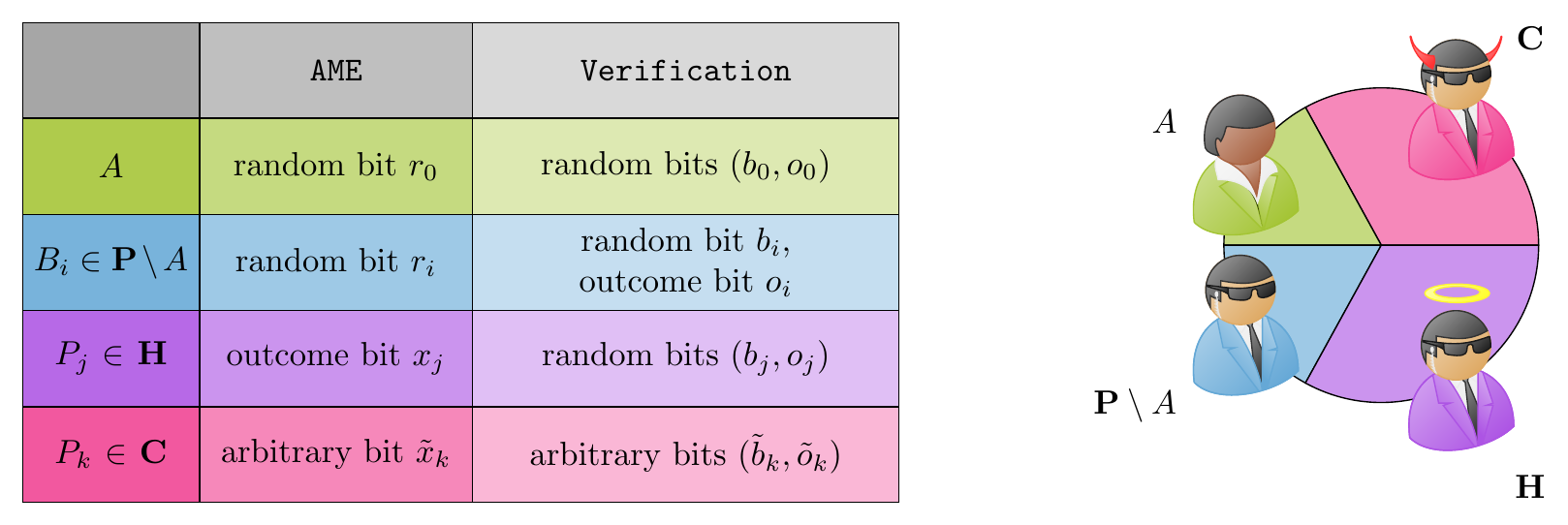}
 \caption{Overview of all public communication for any party in $\net \defeq \aut \discup \uhon \discup \col$ when running the \texttt{AME} and \texttt{Verification} protocols. The communication summarized in the two columns needs to be indistinguishable from the perspective of any Eve. Since $\Alice$ only announces uniformly random and uncorrelated bits, all other communication must follow the same probability distribution. Only the communication from $\col$ can in principle diverge -- should they choose not to hide their identities.}
 \label{tab:publiccommunication}
\end{table*}

\subsection{Anonymity during the \texttt{AME} protocol}\label{app:AME}
\noindent At the start of the \texttt{AME} protocol, the shared quantum state is as given by the following equation:
\begin{equation}
\ket{\net} \approx_{\epsilon} \frac{1}{\sqrt{2}} \left(\ket{0 \hdots 0}_{\hon} \otimes \ket{\Psi}_{\col} + \ket{1\hdots 1}_{\hon} \otimes \ket{\Phi}_{\col}\right).
\label{eq:GHZapp}
\end{equation}
While the \texttt{AME} protocol prescribes measurements to both $\uhon$ and $\ucol$, the parties in $\ucol$ might not measure and announce something unrelated to their arbitrary actions on the quantum state -- therefore we now only calculate the probability of the measurement outcomes $\mu^{\alpha}_{\uhon} = \{\mu_{j} \mid j \in \uhon\}$ of $\uhon$ taking values $x_{\uhon}^{\alpha} = \{x^{\alpha}_{i}\} \in \{0,1\}^\abs{\uhon}$. We want to show that they are uniformly random and that there are no correlations between the outcomes and any Eve that she might exploit, where Eve might be anyone in the network but Alice. That is, we want to show 

\begin{equation}
\Pr\left(\mu^{\alpha}_{\uhon} = x_{\uhon}^{\alpha}\mid \mathcal{I}^{+}_{\mathrm{Eve}},\mathcal{I}_{\mathrm{Eve}}\right) = \Pr\left(\mu^{\alpha}_{\uhon} = x_{\uhon}^{\alpha}\right) = \frac{1}{2^{\abs{\uhon}}},
\label{eq:AMEprivacy}
\end{equation}
where the second equality implies that the probability distribution of the measurement outcomes is uniform and the first equality implies that there are no correlations between the information accessible to Eve -- including her quantum state -- and the measurement outcomes. Moreover, we also want to show that the post-measurement state does not possess any other correlations regarding the roles of the parties that are accessible or exploitable by Eve.

\noindent The measurements on $\uhon$ in the \texttt{AME} protocol are a PVM with outcomes $\{x^{\alpha}_{\uhon}\}$ and associated projectors

\begin{equation}
X^{\alpha}_{\uhon} \defeq H_{\uhon}\ketbra{x^{\alpha}_{\uhon}}{x^{\alpha}_{\uhon}}_{\uhon}H_{\uhon} = \bigotimes_{j \in \uhon} H_{j}\ketbra{x^{\alpha}_{j}}{x^{\alpha}_{j}}_{j}H_{j},
\end{equation}
which results in the probability of the measurement outcome $\mu^{\alpha}_{\uhon}$ taking the value  $x_{\uhon}^{\alpha}$ being given by
\begin{equation}
\begin{split}
\Pr(\mu^{\alpha}_{\uhon} = x_{\uhon}^{\alpha}) &= \tr\big[X^{\alpha}_{\uhon} \ketbra{\net}{\net}\big] \\
&= \frac{1}{2} \tr\big[\left(\ketbra{0 \hdots 0}{0 \hdots 0}_{\aut} \right)\big]\tr\big[X^{\alpha}_{\uhon} \ketbra{0 \hdots 0}{0 \hdots 0}_{\uhon} \big]\tr\big[ \ketbra{\Psi}{\Psi}_{\col}\big] \\
&+ \frac{1}{2} \tr\big[\left(\ketbra{0 \hdots 0}{1 \hdots 1}_{\aut} \right)\big]\tr\big[X^{\alpha}_{\uhon} \ketbra{0 \hdots 0}{1 \hdots 1}_{\uhon} \big]\tr\big[ \ketbra{\Psi}{\Phi}_{\col}\big] \\
&+ \frac{1}{2} \tr\big[\left(\ketbra{1 \hdots 1}{0 \hdots 0}_{\aut} \right)\big]\tr\big[X^{\alpha}_{\uhon} \ketbra{1 \hdots 1}{0 \hdots 0}_{\uhon} \big]\tr\big[ \ketbra{\Phi}{\Psi}_{\col}\big] \\
&+ \frac{1}{2} \tr\big[\left(\ketbra{1 \hdots 1}{1 \hdots 1}_{\aut} \right)\big]\tr\big[X^{\alpha}_{\uhon} \ketbra{1 \hdots 1}{1 \hdots 1}_{\uhon} \big]\tr\big[ \ketbra{\Phi}{\Phi}_{\col}\big] \\
&= \frac{1}{2} \tr\big[ \left(\bigotimes_{j \in \uhon} H_{j}\ketbra{x^{\alpha}_{j}}{x^{\alpha}_{j}}_{j}H_{j} \right) \ketbra{0 \hdots 0}{0 \hdots 0}_{\uhon} \big] \\
&+ \frac{1}{2} \tr\big[\left(\bigotimes_{j \in \uhon} H_{j}\ketbra{x^{\alpha}_{j}}_{j}H_{j} \right) \ketbra{1 \hdots 1}{1 \hdots 1}_{\uhon} \big] \\
&= \frac{1}{2} \left(\prod_{i \in \uhon} \abs{\bra{x^{\alpha}_{i}}\ket{+}}^{2} + \prod_{i \in \uhon} \abs{\bra{x^{\alpha}_{i}}\ket{-}}^{2}\right) \\
&= \frac{1}{2} \left(\frac{1}{2^{\abs{\uhon}}} + \frac{1}{2^{\abs{\uhon}}} \right) = \frac{1}{2^{\abs{\uhon}}}.\\
\end{split}
\end{equation}
This satisfies the second equality in Eq.~\eqref{eq:AMEprivacy}, showing that the measurement outcomes are uniformly random, thereby ensuring that all the communication of the \texttt{AME} column of Tab.~\ref{tab:publiccommunication} is indistinguishable -- excluding the trivial case where $\col$ reveals itself.

\noindent The global post-measurement state $\rho_{\mathrm{post\texttt{AME}}}$ is then
\begin{equation}
\begin{split}
\rho_{\mathrm{post\texttt{AME}}} &= X^{\alpha}_{\uhon} \ketbra{\net}{\net}X^{\alpha}_{\uhon} \\
&= \frac{1}{2} \left(\ketbra{0 \hdots 0}{0 \hdots 0}_{\aut} \right) \otimes X^{\alpha}_{\uhon} \ketbra{0 \hdots 0}{0 \hdots 0}_{\uhon} X^{\alpha}_{\uhon} \otimes \ketbra{\Psi}{\Psi}_{\ucol} \\
&+ \frac{1}{2} \left(\ketbra{0 \hdots 0}{1 \hdots 1}_{\aut} \right) \otimes X^{\alpha}_{\uhon} \ketbra{0 \hdots 0}{1 \hdots 1}_{\uhon} X^{\alpha}_{\uhon} \otimes \ketbra{\Psi}{\Phi}_{\ucol} \\
&+ \frac{1}{2} \left(\ketbra{1 \hdots 1}{0 \hdots 0}_{\aut} \right) \otimes X^{\alpha}_{\uhon} \ketbra{1 \hdots 1}{0 \hdots 0}_{\uhon} X^{\alpha}_{\uhon} \otimes \ketbra{\Phi}{\Psi}_{\ucol} \\
&+ \frac{1}{2} \left(\ketbra{1 \hdots 1}{1 \hdots 1}_{\aut} \right) \otimes X^{\alpha}_{\uhon} \ketbra{1 \hdots 1}{1 \hdots 1}_{\uhon} X^{\alpha}_{\uhon} \otimes \ketbra{\Phi}{\Phi}_{\ucol} \\
&= \frac{1}{2} \left(\ketbra{0 \hdots 0}{0 \hdots 0}_{\aut} \right) \otimes  \ketbra{\uhon}{\uhon} \otimes \ketbra{\Psi}{\Psi}_{\ucol} \\
&+ \frac{1}{2} \left(\ketbra{0 \hdots 0}{1 \hdots 1}_{\aut} \right) \otimes (-1)^{\Delta(x^{\alpha}_{\uhon})}\ketbra{\uhon}{\uhon} \otimes \ketbra{\Psi}{\Phi}_{\ucol} \\
&+ \frac{1}{2} \left(\ketbra{1 \hdots 1}{0 \hdots 0}_{\aut} \right) \otimes (-1)^{\Delta(x^{\alpha}_{\uhon})}\ketbra{\uhon}{\uhon} \otimes \ketbra{\Phi}{\Psi}_{\ucol} \\
&+ \frac{1}{2} \left(\ketbra{1 \hdots 1}{1 \hdots 1}_{\aut} \right) \otimes \ketbra{\uhon}{\uhon} \otimes \ketbra{\Phi}{\Phi}_{\ucol} \\
&= \ketbra{\net_{\mathrm{post\texttt{AME}}}}{\net_{\mathrm{post\texttt{AME}}}},
\end{split}
\end{equation}
where $\ketbra{\net_{\mathrm{post\texttt{AME}}}}$ is the pure state
\begin{equation}
\ket{\net_{\mathrm{post\texttt{AME}}}} = 
    \frac{1}{\sqrt{2}} \left(\ket{0 \hdots 0}_{\aut}   \otimes \ket{\Psi}_{\ucol} +  (-1)^{\Delta(x^{\alpha}_{\uhon})}\ket{1 \hdots 1}_{\aut}   \otimes \ket{\Phi}_{\ucol} \right)\otimes  \ket{\uhon},\label{eq:AMEpostmeasstate}
\end{equation}
showing that the only correlation between the measurement outcome and the state on $\ahon \discup \col$ is in the phase, where one could in principle learn the parity of the measurement outcome $x^{\alpha}_{\uhon}$. However, any such phase estimation is impossible if one does not have access to the complete state (i.e.~tracing out $\ahon$ that does not collude with Eve results in a state on $\col$ that is uncorrelated with the measurement outcome $x^{\alpha}_{\uhon}$). This means that the post-measurement state of any attacker in $\Bobs$ or $\col$ is uncorrelated from the measurement outcome $x^{\alpha}_{\uhon}$ and the roles of $\uhon$. Therefore, for either of these types of Eve everyone in $\uhon$ remains anonymous (cf.~\ame{1} in Tab.~\ref{tab:perspectivefrom}).

Furthermore $\uhon$ is disentangled from the rest of the network and $\ket{\uhon}$ itself is separable over the constituents of $\uhon$. Therefore, nobody in $\uhon$ can learn anything about the roles of any other party in the network. We can conclude that for $\mathrm{Eve}$ in $\uhon$, Def.~\eqref{def:anonymity} holds for any of the subsets of $\net$ (cf.~\ame{2} in Tab.~\ref{tab:perspectivefrom}).

When Eve is a party in $\Bobs$, the roles of the parties in either $\aut$ or $\col$ are hidden because the relevant correlations of the state are unchanged by running the \texttt{AME} protocol -- they essentially share a \GHZ{} state, possibly including some additional phase, and therefore there are no revealing correlations available to anyone but Alice, meaning that here Def.~\eqref{def:anonymity} also holds. The exact same argument holds for Eve in $\col$ with respect to the anonymity of $\aut$ (cf.~\ame{3} in Tab.~\ref{tab:perspectivefrom}).

\subsection{Anonymity during the \texttt{Verification} rounds}\label{app:Ver}
\noindent At the start of the \texttt{Verification} round, the state is the post-measurement state from Eq.~\eqref{eq:AMEpostmeasstate}, up to the correction by $\Alice$. We allow for a faulty correction, therefore keeping the phase arbitrary in the following analysis, writing $(-1)^{\Delta} = \pm 1$ for the phase. We again calculate the probability that, based on some basis choice $\{b_{i}\}$ and given the \texttt{AME} measurement outcome $x^{\alpha}_{\uhon}$, the measurement outcome $\mu^{\alpha} = \{\mu_{j}\mid j \in \Bobs\}$ takes some particular value $o^{\alpha} = \{o^{\alpha}_{i}\} \in \{0,1\}^{\abs{\Bobs}}$, show that the outcome is uniformly random and that there are no correlations between the outcome and the quantum states of all possible Eves. That is, we want to show that
\begin{equation}
\Pr\left(\mu^{\alpha} = o^{\alpha}\mid \mathcal{I}^{+}_{\mathrm{Eve}}, \mathcal{I}_{\mathrm{Eve}}\right) = \Pr\left(\mu^{\alpha} = o^{\alpha}\right) = \frac{1}{2^{\abs{\Bobs}}},
\label{eq:VERprivacy}
\end{equation}
where $\mathrm{Eve}$ may be anyone in $\Bobs$, $\uhon$ or $\col$. Again, we also show that the post-measurement states do not possess any other correlations regarding the roles of the parties which are exploitable by anyone in $\Bobs$, $\uhon$ or $\col$.

Each measurement outcome is associated with a certain measurement projector $O_{\Bobs}^{\alpha}$, which is itself dependent on the basis choice $\{b_{i}\}$. Explicitly, we define

\begin{equation}
O_{\Bobs}^{\alpha}(\{b_{i}\}) \defeq \left(\bigotimes_{\{i \in \Bobs\mid b_{i} = 0\}} H_{i}\ketbra{o^{\alpha}_{i}}{o^{\alpha}_{i}}H_{i} \right)\otimes \left(\bigotimes_{\{i \in \Bobs\mid b_{i} = 1\}} \sqrt{Z_{i}}H_{i}\ketbra{o^{\alpha}_{i}}{o^{\alpha}_{i}}H_{i}\sqrt{Z_{i}}^{\dagger}\right).
\end{equation}
Hence, for any outcome $x^{\alpha}_{\uhon}$ during the \texttt{AME} protocol, the probability of the measurement outcome $\mu^{\alpha}$ being equal to $o^{\alpha}$ becomes (remember that $\Delta$ may depend on $x^{\alpha}_{\uhon}$)
\begin{equation}
\begin{split}
\Pr\left(\mu^{\alpha} = m^{\alpha}\right) = & \tr\big[O^{\alpha}\ketbra{\net_{\mathrm{post\texttt{AME}}}}{\net_{\mathrm{post\texttt{AME}}}}\big] \\
= & \frac{1}{2} \tr\big[\ketbra{0}{0}_{\Alice}\big]\tr\big[O^{\alpha}\ketbra{0 \hdots 0}{0 \hdots 0}_{\Bobs} \big]\tr\big[\ketbra{\uhon}{\uhon}\big] \otimes \ketbra{\Psi}{\Psi}_{\col} \\
+(-1)^{\Delta}& \frac{1}{2} \tr\big[\ketbra{0}{1}_{\Alice}\big]\tr\big[O^{\alpha}\ketbra{0 \hdots 0}{1 \hdots 1}_{\Bobs} \big] \tr\big[\ketbra{\uhon}{\uhon}\big] \otimes \ketbra{\Psi}{\Phi}_{\col} \\
+(-1)^{\Delta}& \frac{1}{2} \tr\big[\ketbra{1}{0}_{\Alice}\big]\tr\big[O^{\alpha}\ketbra{1 \hdots 1}{0 \hdots 0}_{\Bobs} \big] \tr\big[\ketbra{\uhon}{\uhon}\big] \otimes \ketbra{\Phi}{\Psi}_{\col} \\
+ & \frac{1}{2} \tr\big[\ketbra{1}{1}_{\Alice}\big]\tr\big[O^{\alpha}\ketbra{1 \hdots 1}{1 \hdots 1}_{\Bobs} \big]\tr\big[\ketbra{\uhon}{\uhon}\big] \otimes \ketbra{\Phi}{\Phi}_{\col} \\
= & \frac{1}{2} \tr\big[O^{\alpha}\ketbra{0 \hdots 0}{0 \hdots 0}_{\Bobs} \big] \\
+ & \frac{1}{2} \tr\big[O^{\alpha}\ketbra{1 \hdots 1}{1 \hdots 1}_{\Bobs} \big].\\
\end{split}
\end{equation}
Substituting $O^{\alpha}$ we obtain

\begin{equation}
\begin{split}
\Pr\left(\mu^{\alpha} = m^{\alpha}\right) &= \frac{1}{2}
\prod_{\{i \in \Bobs\mid b_{i} = 0\}}
\mel{o^{\alpha}_{i}}{ H_{i}}{0}
\mel{0}{H_{i}}{o^{\alpha}_{i}}
\prod_{\{i \in \Bobs\mid b_{i} = 1\}}
\mel{o^{\alpha}_{i}}{H_{i}\sqrt{Z_{i}}^{\dagger}}{0}
\mel{0}{\sqrt{Z_{i}} H_{i}}{o^{\alpha}_{i}} 
\\
&+ \frac{1}{2}
\prod_{\{i \in \Bobs\mid b_{i} = 0\}}
\mel{o^{\alpha}_{i}}{ H_{i}}{1}
\mel{1}{H_{i}}{o^{\alpha}_{i}} 
\prod_{\{i \in \Bobs\mid b_{i} = 1\}}
\mel{o^{\alpha}_{i}}{H_{i}\sqrt{Z_{i}}^{\dagger}}{1}
\mel{1}{\sqrt{Z_{i}}H_{i}}{o^{\alpha}_{i}},\\
&= \frac{1}{2}\prod_{\{i \in \Bobs\mid b_{i} = 0\}}\abs{\bra{o^{\alpha}_{i}}\ket{+}}^{2}\prod_{\{i \in \Bobs\mid b_{i} = 1\}}\abs{\bra{o^{\alpha}_{i}}\ket{+}}^{2} \\
&+ \frac{1}{2}\prod_{\{i \in \Bobs\mid b_{i} = 0\}}\abs{\bra{o^{\alpha}_{i}}\ket{-}}^{2}\prod_{\{i \in \Bobs\mid b_{i} = 1\}}\abs{\bra{o^{\alpha}_{i}}\ket{-}}^{2} \\
&= \frac{1}{2^{\abs{\Bobs}}},
\end{split}
\end{equation}

\noindent which satisfies the second equation in Eq.~\eqref{eq:VERprivacy}. The global post-measurement state $\rho_{\mathrm{post\texttt{VER}}}$ becomes
\begin{equation}
\begin{split}
 \rho_{\mathrm{post\texttt{VER}}} =& O^{\alpha}\ketbra{\net_{\mathrm{postAME}}}{\net_{\mathrm{postAME}}}O^{\alpha} \\
= &\frac{1}{2} \ketbra{0}{0}_{\Alice} \otimes \big(O^{\alpha}\ketbra{0 \hdots 0}{0 \hdots 0}_{\Bobs} O^{\alpha}\big) \otimes \ketbra{\uhon}{\uhon} \otimes \ketbra{\Psi}{\Psi}_{\col}\\
+(-1)^{\Delta}& \frac{1}{2} \ketbra{0}{1}_{\Alice} \otimes \big(O^{\alpha}\ketbra{0 \hdots 0}{1 \hdots 1}_{\Bobs} O^{\alpha}\big) \otimes  \ketbra{\uhon}{\uhon} \otimes \ketbra{\Psi}{\Phi}_{\col}\\
+(-1)^{\Delta}& \frac{1}{2} \ketbra{1}{0}_{\Alice} \otimes \big(O^{\alpha}\ketbra{1 \hdots 1}{0 \hdots 0}_{\Bobs} O^{\alpha}\big) \otimes  \ketbra{\uhon}{\uhon} \otimes \ketbra{\Phi}{\Psi}_{\col}\\
+ & \frac{1}{2} \ketbra{1}{1}_{\Alice} \otimes \big(O^{\alpha}\ketbra{1 \hdots 1}{1 \hdots 1}_{\Bobs} O^{\alpha}\big) \otimes \ketbra{\uhon}{\uhon} \otimes \ketbra{\Phi}{\Phi}_{\col}\\
= & \frac{1}{2} \ketbra{0}{0}_{\Alice} \otimes \ketbra{\Bobs}{\Bobs}  \otimes \ketbra{\uhon}{\uhon} \otimes \ketbra{\Psi}{\Psi}_{\col} \\
+ \gamma^{\dagger}& \frac{1}{2} \ketbra{0}{1}_{\Alice} \otimes \ketbra{\Bobs}{\Bobs} \otimes  \ketbra{\uhon}{\uhon} \otimes \ketbra{\Psi}{\Phi}_{\col}\\
+\gamma & \frac{1}{2} \ketbra{1}{0}_{\Alice} \otimes \ketbra{\Bobs}{\Bobs} \otimes  \ketbra{\uhon}{\uhon} \otimes \ketbra{\Phi}{\Psi}_{\col}\\
+ & \frac{1}{2} \ketbra{1}{1}_{\Alice} \otimes \ketbra{\Bobs}{\Bobs} \otimes \ketbra{\uhon}{\uhon} \otimes \ketbra{\Phi}{\Phi}_{\col}\big]\\
= & \ketbra{\net_{\mathrm{post\texttt{VER}}}}{\net_{\mathrm{post\texttt{VER}}}},
\end{split}
\end{equation}
where $\gamma=(-1)^{\Delta}\times (-i)^{\abs{\{b_{i}\}}}$ and $\ket{\net_{\mathrm{post\texttt{VER}}}}$ is the pure state 
\begin{equation}
    \begin{split}
        \ket{\net_{\mathrm{post\texttt{VER}}}} \defeq \left(\ket{0}_{\Alice} \otimes \ket{\Psi}_{\col} + \gamma \ket{1}_{\Alice} \otimes \ket{\Phi}_{\col}\right) \otimes \ket{\Bobs} \otimes \ket{\uhon}
    \end{split}
\end{equation}
and $\ket{\Bobs}$ is the state associated with the measurement outcome $o^{\alpha}$
\begin{equation}
    \begin{split}
        \ket{\Bobs} \defeq \left(\bigotimes_{i \in \{\Bobs | b_{i} = 0\}}H_{i}\ket{o^{\alpha}_{i}}_{i}\right) \otimes  \left(\bigotimes_{i \in \{\Bobs | b_{i} = 1\}}\sqrt{Z_{i}}H_{i}\ket{o^{\alpha}_{i}}_{i}\right).
    \end{split}
\end{equation}
From the perspective of $\uhon$, all communication is indistinguishable (cf.~ the \texttt{Verification} column in Tab.~\ref{tab:publiccommunication}); $\uhon$ is   dis-entangled from everyone else and the state on $\uhon$ is itself separable. We can conclude that -- with anyone in $\uhon$ as Eve -- the anonymity of everyone in the network is preserved (cf.~\ver{1} in Tab.~\ref{tab:perspectivefrom}). 

Moreover, $\Bobs$ is dis-entangled from all other parties in the network and their post-measurement state is separable as well. Again, all communication from their perspective is uniformly random (cf.~ the \texttt{Verification} column in Tab.~\ref{tab:publiccommunication}), so we can conclude that -- with anyone in $\Bobs$ as Eve -- the anonymity of everyone in the network is maintained (cf.~\ver{2} in Tab.~\ref{tab:perspectivefrom}). 

The only relevant information is $\abs{\{b_{i}\}}$, which is encoded into the phase of the state on $\Alice \cup \col$; any phase estimation algorithm to retrieve this information would require access to the entire state, including the state of $\Alice$, which is inaccessible to $\col$. Again, from the perspective of $\col$ all communication is indistinguishable (cf.~ the \texttt{Verification} column in Tab.~\ref{tab:publiccommunication}) and we can conclude that -- with $\col$ as Eve -- here too the anonymity of all parties in the network is preserved (cf.~\ver{3} in Tab.~\ref{tab:perspectivefrom}). 

Note that the \texttt{Verification} round can only pass if $\ket{\Psi}_{\col} = \ket{\Phi}_{\col}$, that is when $\col$ is not entangled to $\Alice$ and $\Bobs$. However, this is not a necessary condition for anonymity, since the identity of Alice is preserved even if the \texttt{Verification} round fails.
There is no information encoded into the state regarding the distribution of $\ahon$ and $\uhon$, nor into the measurement outcome $o^{\alpha}$. The only valuable information in the state is the parity of the number of $Y$-measurements, encoded in the phase of the qubit of $\Alice$, which is dis-entangled from all other parties and therefore only accessible to  $\Alice$.

\subsection{Anonymity during the \texttt{KeyGen} rounds}\label{app:key}
\noindent As the \texttt{Verification} rounds ensure that the \GHZ{m+1} state on $\aut$ is dis-entangled from the non-participating parties in $\una$ and after running the \texttt{AME} protocol no party in $\uhon$ is entangled to any other party, all subsets listed in Tab.~\ref{tab:perspectivefrom} are dis-entangled from each other. Hence, we can write the full-network state at the start of the \texttt{KeyGen} round as
\begin{equation}
    \ket{\net_{\texttt{KeyGen}}} \hat{=} \ket{\mathrm{GHZ}}_{\aut} \otimes \ket{\uhon} \otimes \ket{\Psi}_{\col}.
\end{equation}
Since there is no communication during the \texttt{KeyGen} rounds, there is no leakage from $\aut, \uhon, \col$ outside the subset itself (cf.~\key{1} in Tab.~\ref{tab:perspectivefrom}).  As $\ket{\uhon}$ is a separable state, the case $\uhon$ is trivial (cf.~\key{2} in Tab.~\ref{tab:perspectivefrom}). Finally, due to its symmetries, the \GHZ{m+1} state cannot reveal who the parties sharing the state are. This ensures that there is no privacy leakage for $\aut$ either (cf.~\key{3} in Tab.~\ref{tab:perspectivefrom}).

\end{document}